\pdfoutput=1
\documentclass{PoS}

\title{The Pion Form Factor at Large Momentum Transfer}

\ShortTitle{Pion Form Factor at Large Momentum Transfer}

\author{\speaker{Pai-hsien Jennifer Hsu} 
        \\
        Yale University\\
        E-mail: \email{pai-hsien.hsu@yale.edu}}

\author{George T.\ Fleming\\
        Yale University\\
        E-mail: \email{George.Fleming@Yale.edu}}

\abstract{We present our calculations of the electromagnetic form
factor of pions. We explore the properties of pion form factor at
momentum transfer larger than previous studies by including more
combinations of source and sink momenta and using more
configurations.We fit our results using vector meson dominance (VMD)
hypothesis.}

\FullConference{The XXV International Symposium on Lattice Field Theory\\
         July 30-4 August 2007\\
         Regensburg, Germany}

\begin{document}

\section{Motivation}
The pion form factor is often considered a good observable for
studying the onset of the perturbative QCD (pQCD)regime in exclusive
processes. There are several reasons: First, the asymptotic forms of
the pion form factor at both large and small $Q^2$ are known. At
large $Q^2$ it scales as
\cite{Brodsky:1973kr,Brodsky:1974vy,Farrar:1979aw,Radyushkin:1977gp,Efremov:1978,Efremov:1978rn,Efremov:1979qk,
Jackson:1977,Lepage:1979zb}

\begin{equation}
\label{eq:large_qsq_scaling}
F_\pi(Q^2) =
\frac{8\pi\alpha_s(Q^2)f_\pi^2}{Q^2} \quad \mathrm{as} \quad Q^2 \to
\infty
\end{equation}
while at small $Q^2$, the pion form factor can be well described by
the Vector Meson Dominance (VMD) Model \cite{Holladay:1955,
Frazer:1959gy, Frazer:1960}

\begin{equation}
\label{eq:vmd_form} F_\pi(Q^2) \approx \frac{1}{1+Q^2 \left/
m_\mathrm{VMD}^2 \right.} \quad \mathrm{for} \quad Q^2 \ll
m_\mathrm{VMD}^2
\end{equation}
Therefore at some $Q^2$ there must be a transition from the VMD
behavior to the large $Q^2$ scaling predicted by pQCD. Since the
pion is the lightest hadron, the transition is expected to occur at
lower $Q^2$ than heavier hadrons, which makes it relatively easier
to probe by both experiments and Lattice QCD (LQCD). Finally, there
is no disconnected diagram on the lattice for the pion form factor.
Thus the calculation is pretty straightforward. Previous LQCD
studies on the pion form factor can be found in
\cite{Bonnet:2004fr,Hashimoto:2005am,Brommel:2005ee,Brommel:2006ww}
and the references therein.

\begin{figure}[h]
    \centering
        \includegraphics[width=0.66\textwidth]{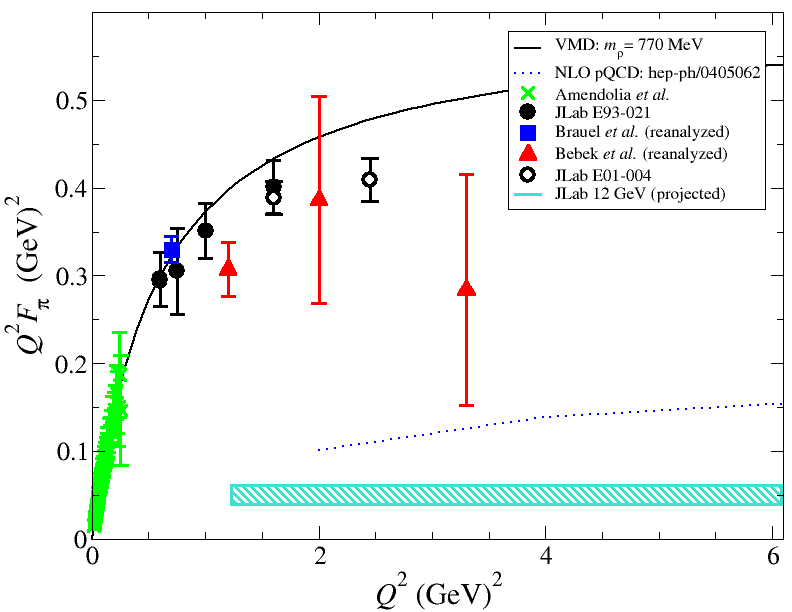}
    \label{fig:Exp}
    \caption{Summary of experimental data
for the pion electromagnetic form factor.  The two points with open
circles are the latest data from the Jefferson Lab (JLab). Shaded
regions are expected sensitivities of future experiments.}
  \end{figure}

The current results from various experiments are shown in Fig.
\ref{fig:Exp}, including the latest results from Jefferson Lab
(JLab) experiments E93-021 \cite{Volmer:2000ek,Tadevosyan:2007} and
E01-004 \cite{Horn:2007}. As we can indicate from the figure, the
data points around $Q^2 \thicksim 2 \mathrm{GeV}^2$ start to show
some hints of deviation from the VMD fit. This is the energy regime
we would like to explore in our study.

\newpage

\section{Lattice Techniques}

\begin{figure}[h]
     \centering
         \includegraphics[width=0.4\textwidth,angle=-90]{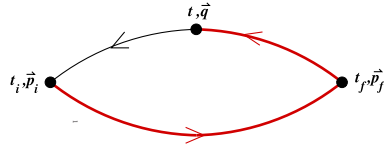}
     \label{fig:ThrPt}
     \caption{The quark propagators used to compute the pion form factor.}
  \end{figure}

In this section we explain the techniques we used in our lattice
calculations, namely the sequential source method (for calculating
the quark propagator) and the ratio method (for the correlation
functions).  The pion electromagnetic form factor is obtained in
LQCD by placing a pion creation operator (the ``source'') at
Euclidean time $t_i$ with momentum $p_i$, a pion annihilation
operator (the ``sink'') at $t_f$ with momentum $p_f$, and a current
insertion at time $t$ with momentum transfer $q$, as shown in Fig.
\ref{fig:ThrPt}. The standard quark propagator calculation provides
the two propagator lines that originate from $t_i$, the remaining
propagator from $t_f$ is obtained by the \textit{sequential source
method}: completely specify the quantum numbers and $p_f$ at the
sink, and contract the propagator from $t_i$ to $t_f$ with the
annihilation operator to serve as the source vector of a second,
sequential propagator inversion. The advantage of using the
sequential source method is that various currents with different
$Q^2$ can be inserted at time $t$ without additional matrix
inversions. The largest $Q^2$ available lies in Breit frame ($\vec
{p_f}= - \vec{p_i}$).

To obtain a simple expression on the lattice, we construct the pion
form factor using the \textit{ratio method}. The pion form factor
$F(Q^2)$ is defined as

\begin{eqnarray}
&&\!\!\!\!\!\!\!\!\!\!\!\!
\left<\pi(\vec{p}_f)\left|V_\mu(0)\right|\pi(\vec{p}_i)\right>_{\rm
continuum}
 \\
&&\!\!\! =
Z_V\left<\pi(\vec{p}_f)\left|V_\mu(0)\right|\pi(\vec{p}_i)\right> =
F(Q^2)(p_i+p_f)_\mu \nonumber
\end{eqnarray}
where $V_\mu(x)$ is the chosen vector current. We can extract
$F(Q^2)$ from some ratio of the three-point correlation function and
the two-point functions.  The three-point function can be written as

\begin{eqnarray}
&&\!\!\!\!\!\!\!\!\!\!\!\!
\Gamma_{\pi\mu\pi}^{AB}(t_i,t,t_f,\vec{p}_i,\vec{p}_f)
 = a^9\sum_{\vec{x}_i,\vec{x}_f}
e^{-i(\vec{x}_f-\vec{x})\cdot\vec{p}_f}
\nonumber \\
&&\times e^{-i(\vec{x}-\vec{x}_i)\cdot\vec{p}_i}
   \left<0\left|\phi_B(x_f)V_\mu(x)\phi_A^\dagger(x_i)\right|0\right>
\end{eqnarray}
where $\phi$'s are operators with pion quantum numbers; $A\in(L,S)$
and $B\in(L,S)$ denote either ``local''($L$) or ``smeared''($S$).
Inserting complete sets of hadron states and requiring $t_i\ll t\ll
t_f$, gives

\begin{eqnarray}
&&\!\!\!\!\!\!\!\!\!\!\!\!
\Gamma_{\pi\mu\pi}^{AB}(t_i,t,t_f,\vec{p}_i,\vec{p}_f) \to
     \left<0\left|\phi_B(x)\right|\pi(\vec{p}_f)\right>
\nonumber \\
&&
     \times\left<\pi(\vec{p}_f)\left|V_\mu(x)\right|\pi(\vec{p}_i)\right>
     \left<\pi(\vec{p}_i)\left|\phi_A^\dagger(x)\right|0\right> \nonumber \\
&&
     \times\frac{a^3}{4E_\pi(\vec{p}_f)E_\pi(\vec{p}_i)} e^{-(t_f-t)E_\pi(\vec{p}_f)}e^{-(t-t_i)E_\pi(\vec{p}_i)}.
\end{eqnarray}
Similarly for the two-point correlator,

\begin{eqnarray}
&&\!\!\!\!\!\!\!\!\!\!\!\! \Gamma_{\pi\pi}^{AB}(t_i,t_f,\vec{p})
 \to \left<0\left|\phi_B(x_i)\right|\pi(\vec{p})\right>
\nonumber \\ && \times
     \left<\pi(\vec{p})\left|\phi_A^\dagger(x_i)\right|0\right>
     \frac{a^3}{2E}e^{-(t_f-t_i)E}.
\end{eqnarray}
We can obtain $F(Q^2)$ from the following ratio

\begin{eqnarray}
F(Q^2) &=& \frac{\Gamma_{\pi
4\pi}^{AB}(t_i,t,t_f,\vec{p}_i,\vec{p}_f)
   \Gamma_{\pi\pi}^{CL}(t_i,t,\vec{p}_f)}
  {\Gamma_{\pi\pi}^{AL}(t_i,t,\vec{p}_i)
   \Gamma_{\pi\pi}^{CB}(t_i,t_f,\vec{p}_f)}
\nonumber \\ && \times
   \left(\frac{2Z_VE_\pi(\vec{p}_f)}{E_\pi(\vec{p}_i)+E_\pi{\vec{p}_f}}\right), \label{theratio}
\end{eqnarray}
where the indices $A$, $B$ and $C$ can be either $L$ (local) or $S$
(smeared).

\section{Simulation Details and Results}

We use lattices generated by MILC \cite{Bernard:2001av}, with volume
$20^3\times32$ and lattice spacing $a=0.125$ fm. The sea quark mass
$m_{sea}$ and the valence quark mass $m_{val}$ are tuned so that we
get the same lightest pion mass $m_\pi(m_{sea})=m_\pi(m_{val})$
\cite{Hagler:2007xi}. The pion operators are fixed at time $t_i=10$
and $t_f=20$, and the number of configurations used in this study is
$201$. We use five different sets of sink momenta: $\vec{p}_f =
(0,0,0), (1,0,0), (1,1,0), (1,1,1)$, and $(2,0,0)$.

\begin{figure}[h]
    \centering
     \includegraphics[width=0.65\textwidth]{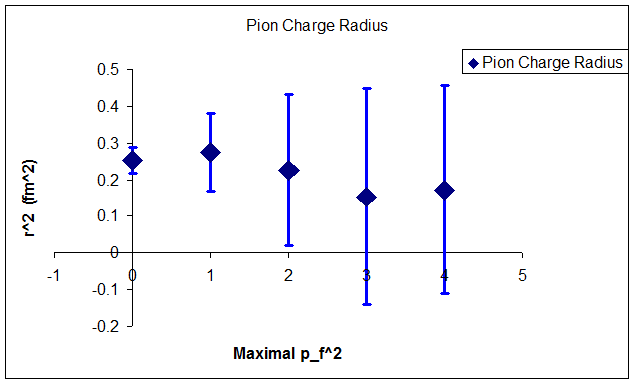}
       \caption{Pion Form Factor VMD fit for $\vec{p}_f = (0, 0, 0)$ to $(2, 0, 0)$.}
     \label{fig:Radius_allPf}
    \end{figure}

We present our results in terms of the square of the pion charge
radius, obtained by the VMD fit:

\begin{equation}
 \langle r_\pi^2 \rangle= \frac{6}{m_{VMD}^2}
\label{eq:charge_radius}
\end{equation}
as shown in Fig. \ref{fig:Radius_allPf}. The first point on the left
is from the data set with only zero sink momentum ($\vec{p}_f = (0,
0, 0)$), and for the next point we combined the data from both
$\vec{p}_f = (0, 0, 0)$ and $\vec{p}_f = (1, 0, 0)$, and for the
third point we added in $\vec{p}_f = (1, 1, 0)$, and so on.

We can see from Fig. \ref{fig:Radius_allPf} that the error bars of
$r_\pi^2$ increases as higher sink momenta are included. Since the
pion charge radius is related to the slope of $F(Q^2)$ \emph{at low
$Q^2$}, we derive $\langle r_\pi^2 \rangle$ from the data set of
zero sink momentum $\vec{p}_f = (0,0,0)$ and $Q^2 < 1
\mathrm{GeV}^2$, and check the consistency between the VMD fit and
the data above $1 \mathrm{GeV}^2$ to see if there is any deviation
from the VMD model.

\begin{figure}[h]
       \centering
         \includegraphics[width=0.75\textwidth]{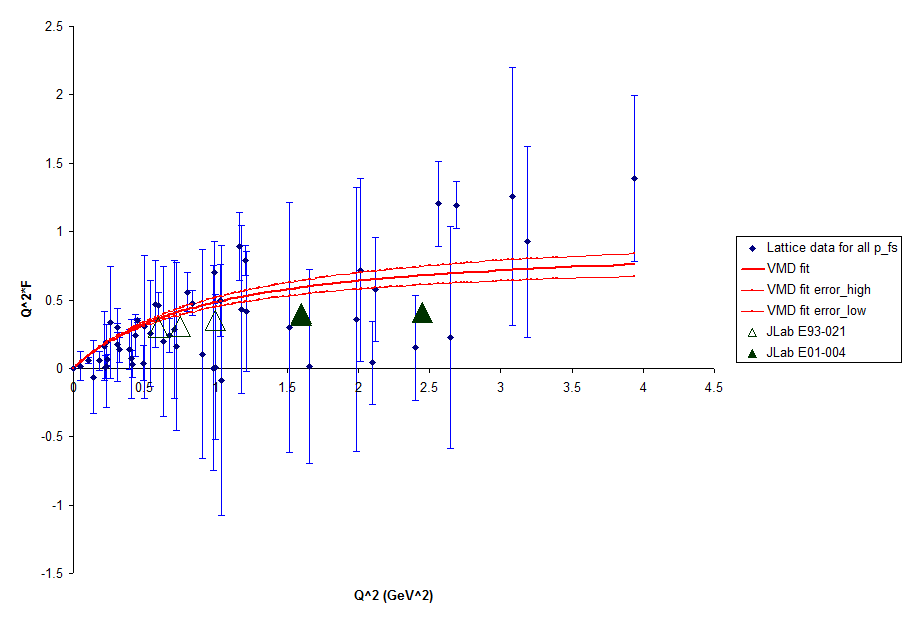}
       \caption{Consistency between data and the VMD fit from low $Q^2$ with zero sink momentum. The three red lines
        represent the VMD fit and its error bars, and the triangle points correspond to the experimental data from JLab.}
       \label{fig:QsqF_with_exp}
   \end{figure}

\begin{figure}[h]
       \centering
         \includegraphics[width=0.65\textwidth]{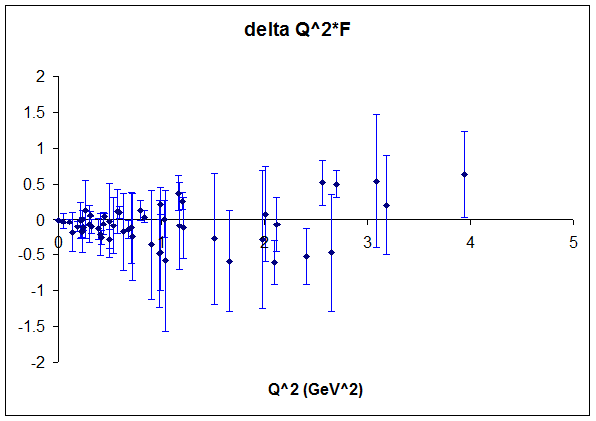}
       \caption{$\Delta Q^2F(Q^2)$, as defined in the text.}
       \label{fig:delta_QsqF}
   \end{figure}

The result of this ``consistency check'' is presented in Fig.
\ref{fig:QsqF_with_exp}, where we plot $Q^2F(Q^2)$ against $Q^2$.
While the quantity $Q^2F(Q^2)$ should approach a constant as
predicted by VMD, we can see that there are some hints of deviation
from the VMD model for points with $Q^2
> 2\mathrm{GeV}^2$.  To further emphasize this observation, we
define $\Delta Q^2F(Q^2)= Q^2F(Q^2)_{Lattice}- Q^2F(Q^2)_{VMD}$, and
plot $\Delta Q^2F(Q^2)$ against $Q^2$ in Fig. \ref{fig:delta_QsqF}.

  \begin{figure}[h]
       \centering
         \includegraphics[width=0.65\textwidth]{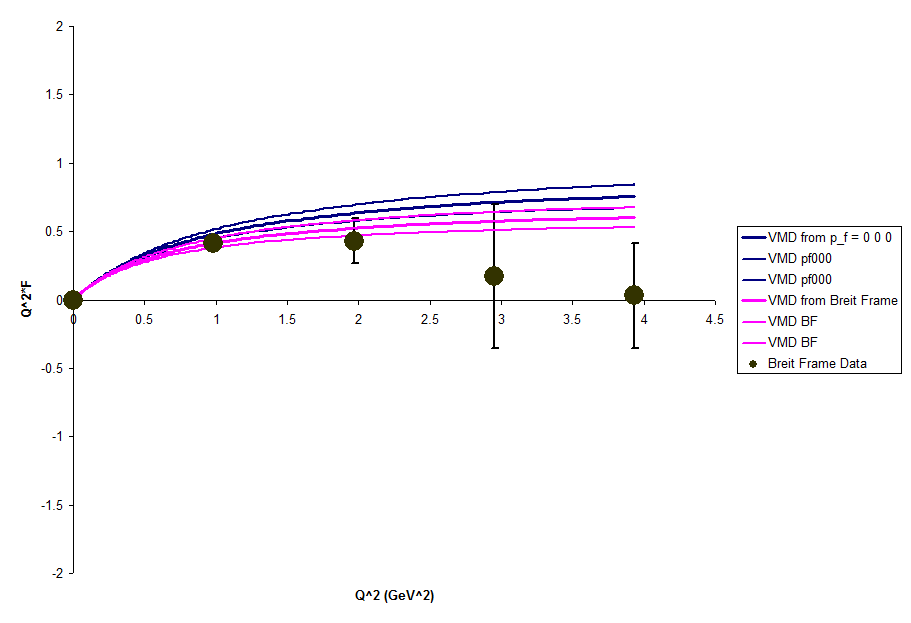}
         \caption{VMD fits with error bands from both data of low $Q^2$ with zero $\vec{p}_f$, and from data points
         in the Breit frame.}
       \label{fig:BF_QsqF_with_VMD}
   \end{figure}

We also compared this VMD fits from low $Q^2$ with that from the
Breit frame (where$\vec {p_f}= - \vec{p_i}$), for the data points of
the Breit frame have relatively small error bars at high $Q^2$. The
result is shown in Fig. \ref{fig:BF_QsqF_with_VMD}. The figure
implies that the VMD fit from the Breit frame (the purple one) is
about $1 \sigma$ away from the fit of a single zero sink momentum
(the blue one), hence exploring further in the Breit frame may be
the correct direction for studying the pion form factor at higher
momentum transfer.


\section{Summary and Outlook}

In this study we have acquired enough lattice data for $Q^2 < 1
\mathrm{GeV}^2$ to extract a reliable pion charge radius $r_\pi$. By
comparing the VMD fit from data with low $Q^2$ and high $Q^2$, we've
started to see some hints of discrepancy between data points at
different momentum transfer, which may indicate the transition from
the VMD model to pQCD, the goal we are seeking for. We also found
that the VMD fit from the Breit frame  is about $1 \sigma$ away from
the fit of a single zero sink momentum, and we infer that we may
explore further in the high $Q^2$ regime by studying the data from
the Breit frame. We are generating four times more data to shrink
the error bars in the Breit frame in the hope of a clearer and
stronger evidence of the transition into the perturbative QCD
regime.

In the meantime, the JLQCD Collaboration has also reported their
calculation of the pion form factor based on all-to-all propagators.
Interested readers may find details in their upcoming publication
\cite{Kaneko:2007}.

\end{document}